\documentclass[twocolumn, twocolappendix]{aastex63}

\usepackage[english]{babel}

\usepackage{amsmath}
\usepackage{graphicx}
\usepackage[version=4]{mhchem}
\usepackage{amsmath}

\usepackage{color}
\definecolor{michiel}{RGB}{51,160,44}

\begin{document}
\title{A potential site for wide-orbit giant planet formation in the IM Lup disk}

\correspondingauthor{Arthur  Bosman}
\email{arbos@umich.edu}
\author[0000-0003-4001-3589]{Arthur D. Bosman}
\affiliation{Department of Astronomy, University of Michigan,
323 West Hall, 1085 S. University Avenue,
Ann Arbor, MI 48109, USA}

\author[0000-0001-6168-1792]{Johan Appelgren}
\affiliation{Lund Observatory, Department of Astronomy and Theoretical Physics, Lund University, Box 43, 22100 Lund, Sweden}

\author[0000-0003-4179-6394]{Edwin A. Bergin}
\affiliation{Department of Astronomy, University of Michigan,
323 West Hall, 1085 S. University Avenue,
Ann Arbor, MI 48109, USA}

\author[0000-0001-9321-5198]{Michiel Lambrechts}
\affiliation{Center for Star and Planet Formation, GLOBE Institute, University of Copenhagen, Øster Voldgade 5-7, 1350 Copenhagen, Denmark }
\affiliation{Lund Observatory, Department of Astronomy and Theoretical Physics, Lund University, Box 43, 22100 Lund, Sweden}

\author[0000-0002-5893-6165]{Anders Johansen}
\affiliation{Center for Star and Planet Formation, GLOBE Institute, University of Copenhagen, Øster Voldgade 5-7, 1350 Copenhagen, Denmark }
\affiliation{Lund Observatory, Department of Astronomy and Theoretical Physics, Lund University, Box 43, 22100 Lund, Sweden}


\begin{abstract}
The radial transport, or drift, of dust has taken a critical role in giant planet formation theory. However, it has been challenging to identify dust drift pile ups in the hard-to-observe inner disk. We find that the IM Lup disk shows evidence that it has been shaped by an episode of dust drift. Using radiative transfer and dust dynamical modeling we study the radial and vertical dust distribution. We find that high dust drift rates exceeding {110} $M_\oplus$ Myr$^{-1}$ are necessary to explain both the dust and CO observations. Furthermore, the bulk of the large dust present in the inner 20 au needs to be vertically extended, implying high turbulence ($\alpha_{z} \gtrsim 10^{-3}$) and small grains (0.2-1 mm). We suggest that this increased level of particle stirring is consistent with the inner dust-rich disk undergoing turbulence triggered by the vertical shear instability. The conditions in the IM Lup disk imply that giant planet formation through pebble accretion is only effective  outside 20 au. If such an early, high turbulence inner region is a natural consequence of high dust drift rates, then this has major implications for understanding the formation regions of giant planets including Jupiter and Saturn.
\end{abstract}






\section{Introduction}

The formation of giant planets, like Jupiter in our own solar system, has long been a challenge for planet formation models. The gas-dominated mass budget of these planets implies an early formation of the planet. This is, within 1-3 Myr of the formation of the host star while abundant gas is still present in the proto-planetary disk \citep[e.g.][]{Haisch2001}. The main bottleneck in the creation of a giant planet is the formation of the core, which has to grow massive enough to start runaway gas accretion \citep[e.g.][]{Pollack1996}.  
Traditional oligarchic models of giant planet core formation models have difficulties with reaching this critical mass within the time-scale necessary for the accumulation of a massive atmosphere{ \citep[see, e.g.][]{Johansen2019}}. This limits the formation of giant planets relatively close to the star (inner few au). 

An alternative method of core formation, pebble accretion, in which a core grows by the rapid accretion of millimeter-centimeter sized solids (pebbles) has slowly been gaining traction \citep[e.g.][]{Johansen2017, Ormel2017review, Drazkowska2022}. This method has the advantage that giant planet cores can be built quickly ($<$1 Myr) and at a far wider range of radii \citep{Lambrechts2012,Bitsch2019, Johansen2019}. However, it requires that millimeter-to-centimeter sized solids (pebbles) to be rapidly transported radially inward through the disk mid-plane, a process know as pebble drift \citep[][]{Weidenschilling1977}. This process has long been proposed to be present in proto-planetary disks, but little to no observational evidence was present. {However, models of efficient drift generally result in a pile-up of dust in the inner disk which should be observable \citep[e.g.][]{Birnstiel2012, Pinte2014}. }

The advent of ALMA has significantly changed the landscape of dust physics and planet formation. High resolution observations are finding structures in many disks at large ($> 10$ au) radii \citep{Andrews2018, Huang2018}that are most likely planet induced.  This implies that massive planets can form quickly at large radii \citep{Zhang2018, Teague2019}. However, other origins for the disk structures have been suggested \citep{Rabago2021, Hu2022}. Furthermore ALMA has been able to confirm the existence of pebbles in a thin layer near the disk mid-plane \citep[e.g.][]{Pinte2016, Villenave2022}. These observations lent strong credence to pebble accretion as a major planet formation pathway. 

Direct evidence of strong pebble drift as a robust physical process that should take place in protoplanetary disks has so far been missing. {It is suggested that the ratio between gas and dust disk size, if suitably large, $R_\mathrm{gas}/R_{\mathrm{dust}} > 3$ could be a signpost of drift. However this is only seen in a small subset of disks \citep[$\sim$ 15\%, e.g.][]{Trapman2019}, indicating that drift happens only in a small fraction of disks, or, more likely, that radial drift of dust can leave a remnant radially-extended dust disk behind. }Further evidence for radial drift can be found in the transport of ices on pebbles surfaces. When the pebbles drift, from the cold outer disk to the warmer inner regions, they bring with them an ice mantle. When the pebble reaches warmer regions, species in the ice sublimate, enriching the gas \citep[e.g.][]{Cuzzi2004}. An enhancement in CO gas has been found around the temperature that CO should sublimate in the HD 163296 disk, which has been attributed to dust drift  \citep{Zhang2020}. Finally, a relation has been found between the water emission coming from the inner most regions, and the sizes of pebble disks, a possible proxy for pebble drift efficiency \citep{Banzatti2020}. 
While these papers all give evidence for drift happening, they do not give a quantitative pebble drift rate, and so can not comment on the effect of drift on planet formation. 

Recent observations of the young ($\sim$ 1 Myr) IM Lup disk have provided more direct evidence \citep{Mawet2012, Cleeves2018, Bosman2021bMAPS, Sierra2021MAPS}. Analysis of the inner $\sim$20 au imply that this region contains a surface density of large dust that is 10--100 times higher than expected from an extrapolation of the outer disk. This is a smoking gun for efficient pebble drift, and would imply pebble drift rates $>$ 40 $M_\oplus$ Myr$^{-1}$. In \citet{Bosman2021bMAPS} we showed that an inner disk pile-up is consistent with the CO isotopologue data. Here we take a deeper look at the total mass concentrated in the inner disk and the implication these observations have on planet formation inside 20 au.

\section{Methods}
\begin{figure}
    \centering
    \includegraphics[width = \hsize]{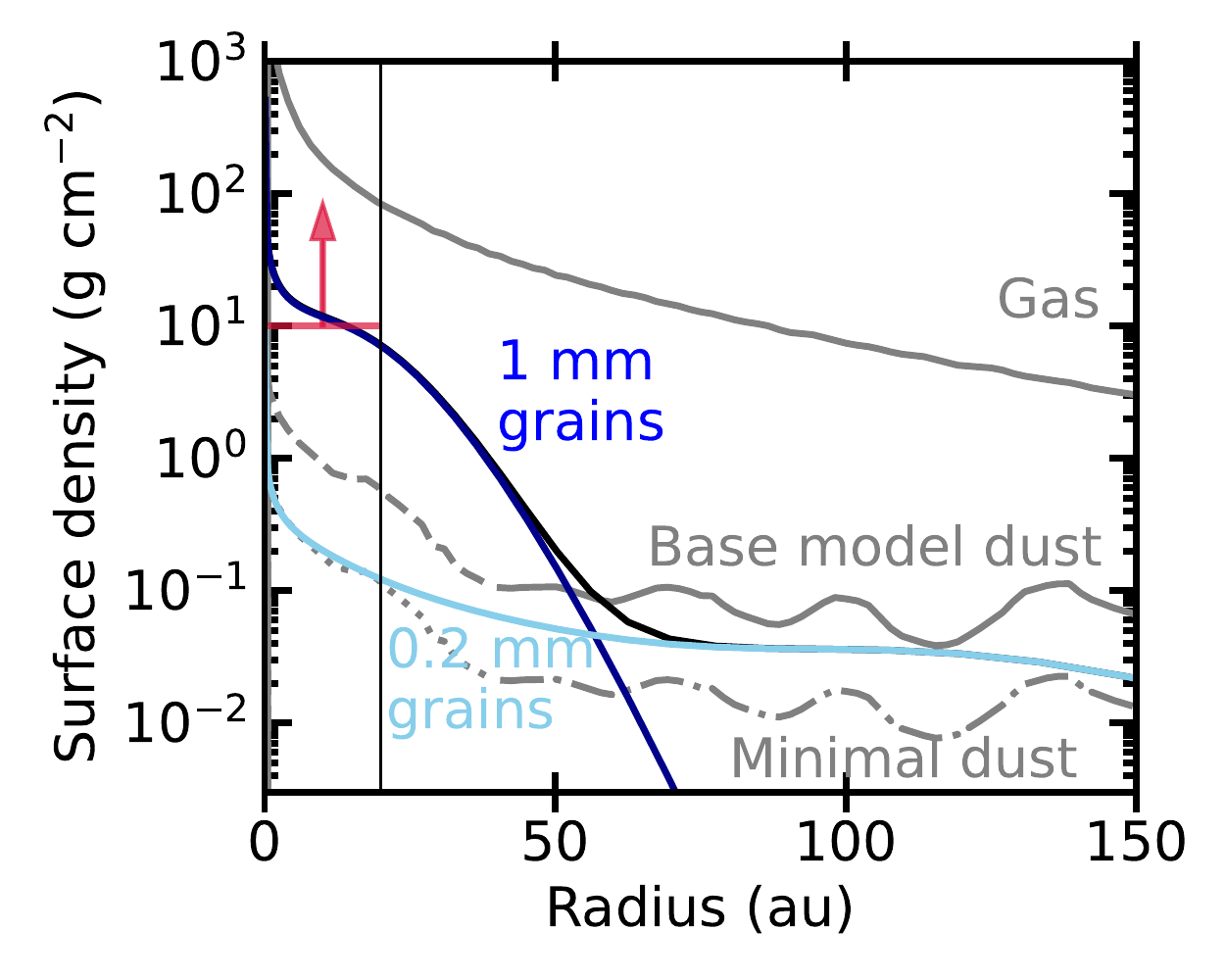}
    \caption{{Surface densities of gas and dust of IM Lup. The solid grey lines show the gas and dust from the model from \citet{Zhang2021MAPS}, which is the baseline for our DALI models. The dash-dotted grey line shows the minimal amount of dust required if purely 0.2 mm particles are responsible for the 1.3 mm dust emission. In the inner 40 au, the dust is optically thick and thus the derived surface density is uncertain.  Dark and light blue lines show the predictions of our drift model for 1 mm, assuming a global gas-to-dust ratio of 100, and 0.2 mm grains, assuming a global gas-to-dust ratio of 1000, respectively (see App.~
    \ref{app:Dust_evolution_model}). Both models are taken at 0.6 Myr, the time at which the 1 mm grains have a large pile-up in the inner 20 au.
    The dark red arrow shows the limit for 400 $M_
    \oplus$ within 20 au, as derived previously \citep{Bosman2021bMAPS}. Our dust model predicts that we can create an inner disk pile-up with significantly mass in the inner disk with 1 mm grains, while just 10\% of the total dust mass in 0.2 mm grains leaves enough opacity in the outer disk to explain the continuum emission further out in the disk.}}
    \label{fig:surface_dens_var}
\end{figure}

To extract the physical conditions in the inner IM Lup disk, we combine a thermo-chemical model \citep{Bruderer2012, Bruderer2013} with a physical setup developed for IM Lup \citep{Zhang2021MAPS}. We then compare our observational findings with a disk evolution model that includes radial dust drift  \citep{Appelgren2020}. From this model we simulate CO isotopologue observations that we compare to observations \citep{Law2021aMAPS}. The base of the model is the gas and dust structure from \citet{Zhang2021MAPS}.  The full model setup is detailed in Appendix~\ref{app:therm_model} and the surface densities used are shown in Fig.~\ref{fig:surface_dens_var}. 
The model contains two dust components, small dust (0.005 $\mu$m--1$\mu$m which follows the gas distribution, and a large dust component (0.005 $
\mu$m--1000$\mu$m) which has a varying scale-height scaling. 

The inner disk of IM Lup shows a flux depression in CO isotopologues \citep[Fig.~\ref{fig:C18O_rad}, grey line;][]{Law2021aMAPS}. Especially \ce{C^{18}O} stood out, as the wings of the \ce{C^{18}O} line profile showed that the flux from the inner 20 au was suppressed below the detection limit. {A fit to the \ce{CO} isotopologue emission with a decrease in the CO abundance required an excessively low CO abundance \citep{Zhang2021MAPS} with the two $J=$2--1 and $J=$1--0 \ce{C^{18}O} line predicting CO columns that are an order of magnitude apart. As such, in our model we assume a constant CO abundance, and instead use dust opacity to lower the inner disk line flux. }
As discussed in \citet{Bosman2021bMAPS}, suppression of line emission requires the large dust grains to be significantly vertically extended, so that the line photons can be scattered and absorbed by the dust. Therefore our models assume that the large dust is fully vertically extended, that is the dust scale-height is the same as the gas scale-height within 30 au.   

{To get a physically motivated dust surface density, a }dust evolution model tuned to the IM Lup disk is run for {1} Myr (see Appendix~\ref{app:Dust_evolution_model} for the full model setup). The resulting dust pile-up is used a physically motivated guide for the radial dust density structure of the inner disk.  For the outer disk we use the observationally derived dust surface density \citep{Zhang2021MAPS}. We take this dust surface density as our base, no drift, dust model. These dust surface densities are not accurate in the inner optically thick regions of the disk. Here we replace the image derived dust surface density by the{ 1 millimeter grain} drift model prediction (Appendix~\ref{app:Dust_evolution_model}). {This model contains $\sim$ 540 M$_\oplus$ within the inner 20 au. We also use inner disk dust surface densities where the surface density is scaled up and down by a factor two to capture the uncertainties in the dust drift models.}

While we use the dust and gas surface density of the \citet{Zhang2021MAPS} model in our models, we do not expect perfect agreement between our model and the data that was fitted by the \citet{Zhang2021MAPS} model. This is due to the vertically extended dust that we include in the inner 30 au in all or our models. This strongly changes the radiation field in the entirety of the disk. This will affect the dust temperature, and thus the continuum flux, as well as strongly affect the gas temperature and thus the line flux of \ce{^{13}CO}, as well as \ce{^{18}CO} at larger radii. A miss match in absolute flux between the model and the observations is thus expected. This should not impact the masses derived for the inner disk.




\section{Results}

\begin{figure}
    \centering

    \includegraphics[width = \hsize]{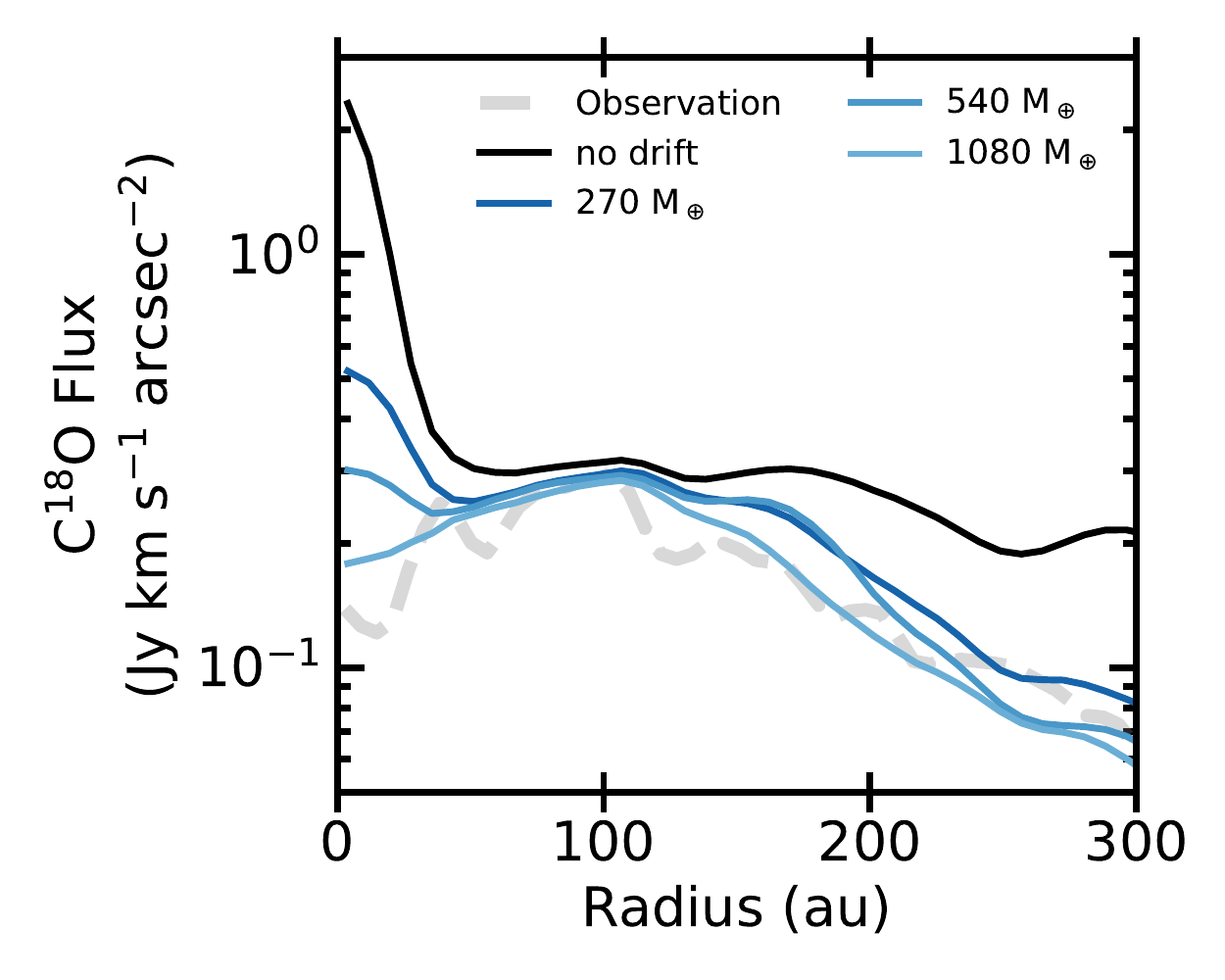}

    \caption{Comparison between the observed \ce{C^{18}O} radial profiles \citep{Law2021aMAPS} (grey) and simulated observations. A model without drift is shown in black. Colored lines show models with an increased surface density in the inner disk due to drift. The 540 $M_\oplus$ model is the 1 mm drift model shown in Fig.~\ref{fig:surface_dens_var}, the others are the same model scale up or down a factor of two in surface density. An inner disk mass of more than 540 $M_\oplus$ within 20 au is required to have an inner disk depression. }
    \label{fig:C18O_rad}
\end{figure}
Figure~\ref{fig:C18O_rad} shows the comparison between the \ce{C^{18}O} model simulation and observations. The base model, without any drift, exhibits a centrally peaked CO emission profile inside 20 au. However, the data requires a strong dip in this region. Such a dip in the inner $\sim$20 au is only present in the simulation if the inner disk dust surface density is at least 10 g cm$^{-2}$, or a gas-to-dust ratio $<$10. At that point the dust $\tau =1$ at 1.3 mm sits around or above the $\tau = 1$ surface for the \ce{C^{18}O} $J=$2--1 line emission, allowing the dust to absorb the line photons. 

This corresponds to a mass reservoir of at least 540 $M_\oplus$ within 20 au. The model with 1080 $M_\oplus$ show behavior closer to that of the observations. The comparison between the model and observations for \ce{^{13}CO} also suggest that higher dust surface densities are required {(Fig.~\ref{fig:13CO_rad})}. However, for our further derivations we will use the lower limit of 540 $M_\oplus$. 

{The dust content in the inner disk expected from the gas surface density and a gas-to-dust ratio of 100  -- without the presence of any dust transport -- is $\sim$ 20 $M_\oplus$. Therefore, most of the mass currently inferred in the inner disk had to be have been transported into the inner 20 au from the outer disk.}  With the current age of IM Lup of $<$ 1 Myr \citep{Mawet2012}, this would require a dust drift rate of $>540$ $M_\oplus$ Myr$^{-1}$ .

\subsection{Dust mass in the inner disk}

\begin{figure*}
    \centering
    \begin{minipage}{0.49\hsize}
    \includegraphics[width = \hsize]{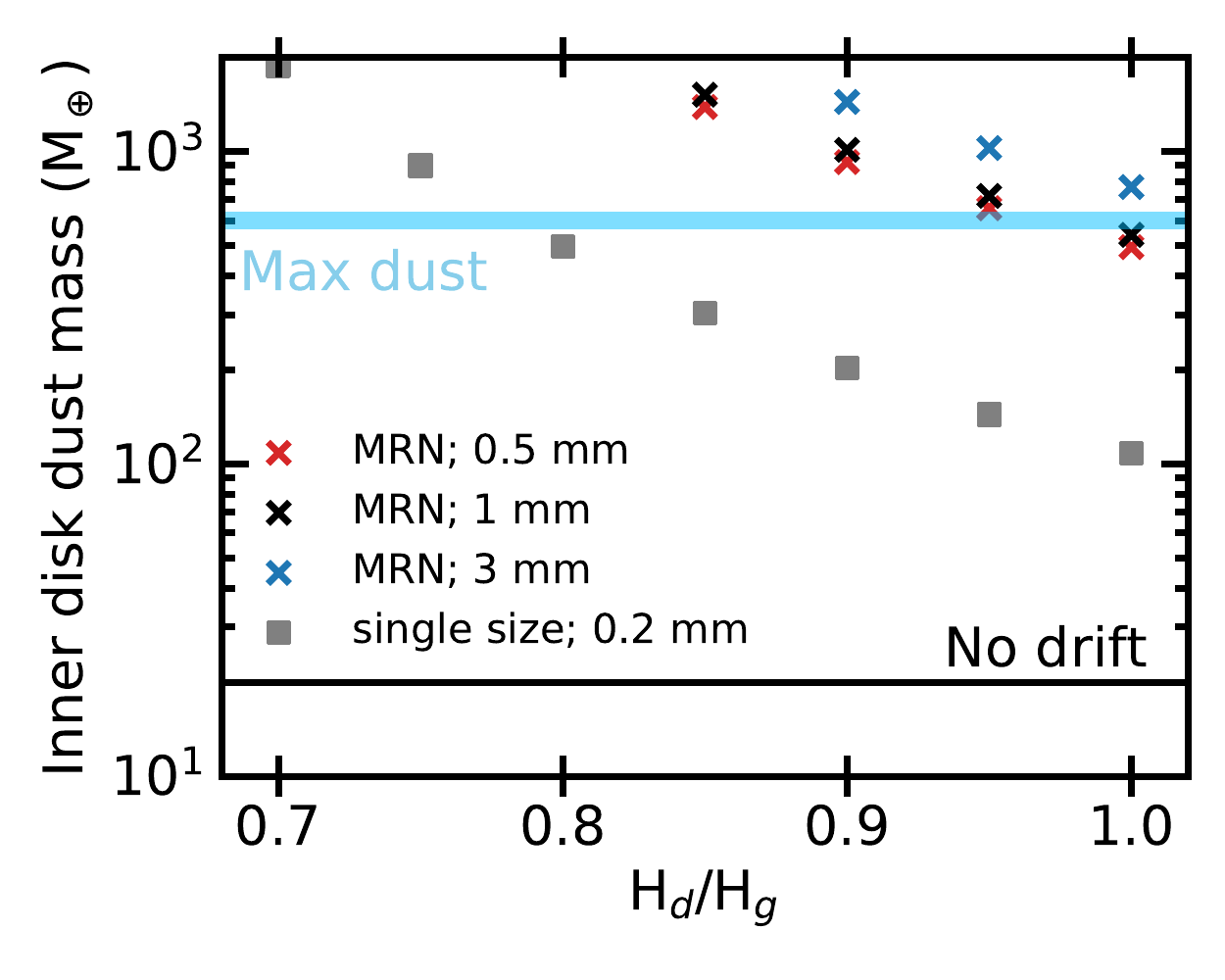}
    \end{minipage}%
    \begin{minipage}{0.49\hsize}
    \includegraphics[width = \hsize]{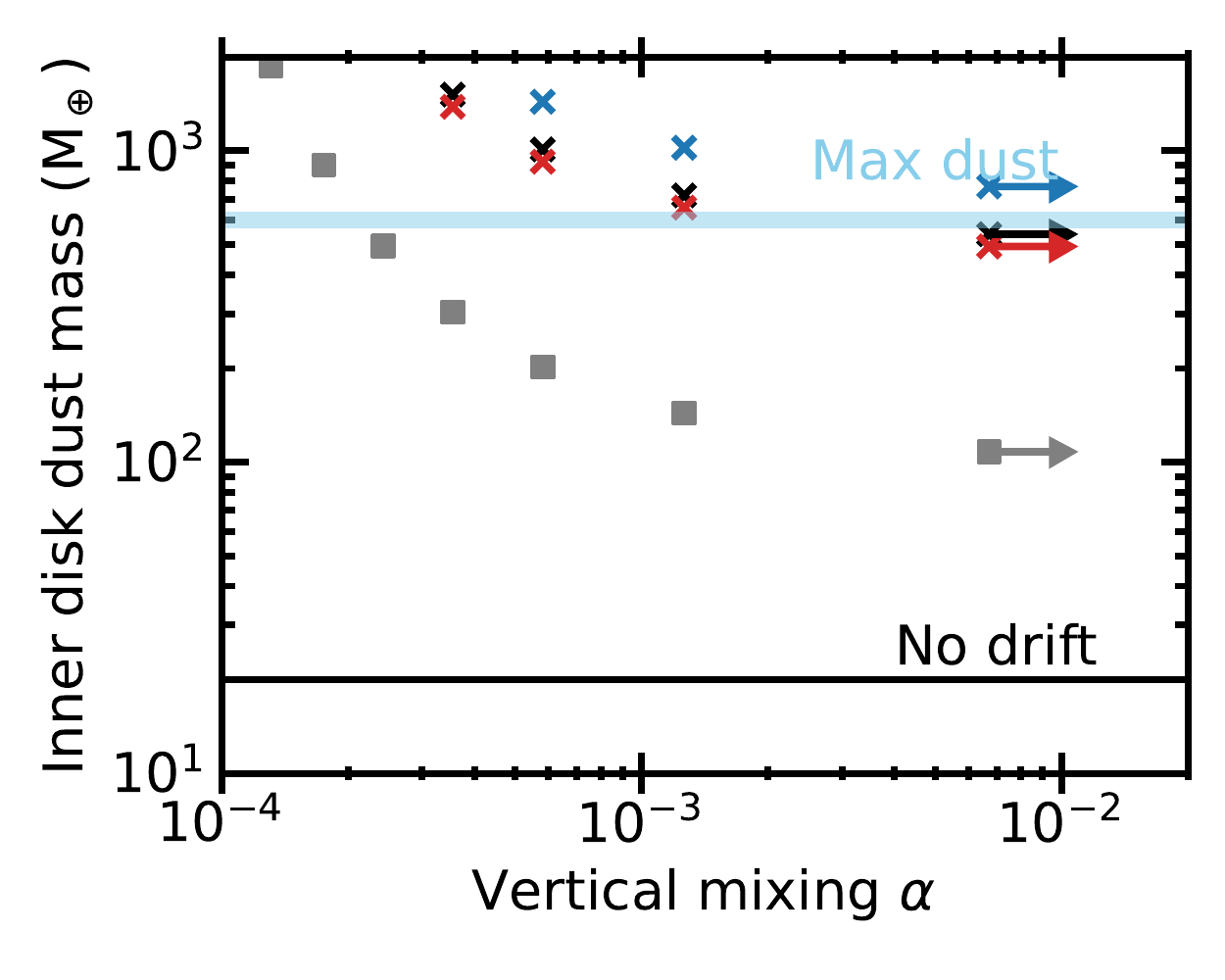}
    \end{minipage}    
    \caption{Left: Required dust mass to suppress the \ce{C^{18}O} in the inner 20 au of IM Lup as function of the dust scale-height. Crosses assume a dust size distribution from 0.005 $\mu$m to the listed size with a powerlaw slope of -3.5. Squares assume all the dust is a single size, 0.2 mm, which yields the highest opacity at 1.3 millimeter. Blue vertical line shows the maximal total refractory mass in the IM Lup disk. Right: Same as on the left, but the H$_d$/H$_g$ converted into a vertical mixing $\alpha$ assuming the dust of 0.2 mm size is dominating the opacity and needs to be elevated to the listed Hd/Hg \citep{Johansen2014}. Points with arrows take the mass for H$_d$/H$_g$ = 1, and show the alpha for H$_d$/H$_g$ = 0.99, as an infinite $\alpha$ is required for H$_d$/H$_g$ = 1. A high $\alpha > 0.001$ is required when a full dust size distribution is assumed. Only under the assumption that the dust is 0.2 mm sized will $\alpha < 10^{-3}$ solutions be possible. }
    \label{fig:Mass_limits}
\end{figure*}

A dust size distribution with a higher opacity would require less mass, while a dust size distribution that is less vertically extended would require more mass. To explore this we calculate the dust mass required for dust distributions. For simplicity we describe the vertical extend of the millimeter-opacity and mass dominating large dust can be modeled with a single $H_d/H_g$. Our standard model uses the ``large grain'' dust opacity from \citet{Birnstiel2018}. This assumes a dust size distribution between 0.005 $\mu$m and 1 mm, with a power-law size distribution with a slope of -3.5. Opacities were also calculated assuming the largest grains are 0.5 or 3 mm as well as under the extreme assumption when all dust is 0.2 mm in size. The latter assumption provides a strict lower limit, as 0.2 mm grains give the maximum extinction per unit dust mass at 1.3 mm. We also calculated the mass required if the dust was more vertically settled, by requiring the dust $\tau=1$ surface at 1.3 mm to be at the same height as in our $H_d/H_g =1$ model. 


Figure \ref{fig:Mass_limits} show the required dust masses under these assumptions. These dust masses are compared to the maximal amount of dust that can be present in the full IM Lup disk. The gas mass in the IM Lup disk model is 0.2 $M_\odot$, about 20\% of the stellar mass \citep{Zhang2021MAPS}. This is on the upper end of the mass the disk can have before it gets dynamically unstable \citep{Toomre1964}. Assuming a standard ISM gas-to-dust ratio of 100, that corresponds to about 600 M$_\oplus$ of refractories, or $\sim 1000$ M$_\oplus$ of total solids assuming all \ce{H2O} and \ce{CO} is frozen out. 
For grains size distributions that include grains larger than 5 mm or single grains size populations smaller than 0.15 mm,{ the required mass to efficiently extinct the CO emission is greater than the total available mass.}

Figure~\ref{fig:Mass_limits} also shows that the large dust must be significantly lofted into the disk surface layers, in the optimal case of pure 0.2 mm grains, the dust scale-height is $>0.75$ times the gas scale height. In the cases that there are larger grains in the inner disk, the many opacity providing grains of $\sim$ 0.2 mm sizes need to be lofted to at least 90\% of the gas scale height. Neither case is close to the expected thin mid-plane layer of grown dust that is seen in the outer regions of disks \citep[e.g.][]{Dullemond2018}.

The extended vertical distribution of grains implies that the dust is being mixed up by some form of turbulence. We can derive a turbulent $\alpha_z$ from the particles Stokes number ($\mathrm{St}$) and the required $H_d/H_g$ using 
\begin{equation}
    \frac{H_d}{H_g} = \sqrt{\frac{\alpha_z}{{\rm St}+\alpha_z}}
\end{equation}
\citep{Johansen2014}. To get a strict lower limit on $
\alpha$, we take the Stokes number of the opacity dominant 0.2 mm grains. 
This reveals that for the grain size distributions, high levels of $\alpha$ are required, $\alpha_z >> 10^{-3}$, to keep the inner disk mass reservoir under the total available disk dust mass. In the extreme case of all dust mass confined to 0.2 mm grains, $\alpha_z$ down to $2 \times 10^{-4}$ is possible, but this scenario is highly unlikely. A grain distribution out to grains of 3 mm seem to be ruled out, on the condition that $\alpha_z \gg 10^{-2}$.

\subsection{Turbulent implications}
The CO isotopologues are implying that $\gg$ $100$\,M$_\oplus$ of dust is piled-up within the inner $20$\,au of the disk. This requires significant dust drift rates $\gg 100$\,M$_\oplus\,\mathrm{Myr}^{-1}$. 
Our dust dynamical models show that enough dust can be transported into the inner disk if dust sizes are around 1 millimeter.
This particle size is consistent with the fragmentation limit in disks with low turbulent stirring, where $\alpha$ is $\lesssim 10^{-3}$\citep{Birnstiel2010}.
Such a low degree of particle turbulence is consistent with dust observations of the outer parts of protoplanetary disks \citep{Pinte2022}.  
In the inner disk however, the dust needs to be efficiently vertically distributed. This requires a high levels of vertical turbulence ($\alpha_z > 10^{-3}$). {The MRI, and other (non-ideal) MHD effects can stir up the dust particles \citep{Riols2018, Yang2018}, however, it is not clear, why this would lead to such a strong discrepancy in particle scale-heights as seen in IM Lup.}



{It is thus more likely that there is another process that increases the vertical turbulence and elevates grains to the heights required by the observations. }
A prime candidate for driving large-scale vertical gas motions is the vertical shear instability \citep[VSI,][]{Urpin1998,Nelson2013}.
{This instability is triggered when the stabilizing vertical buoyancy forces are overcome by vertical shear in the disk, which requires gas cooling timescales that are shorter than orbital timescales, $t_{\rm cool} \ll t_{\rm orbit}$ \citep{Lin2015, Fukuhara2021}. }

\begin{figure}
    \centering
    \includegraphics[width = \hsize]{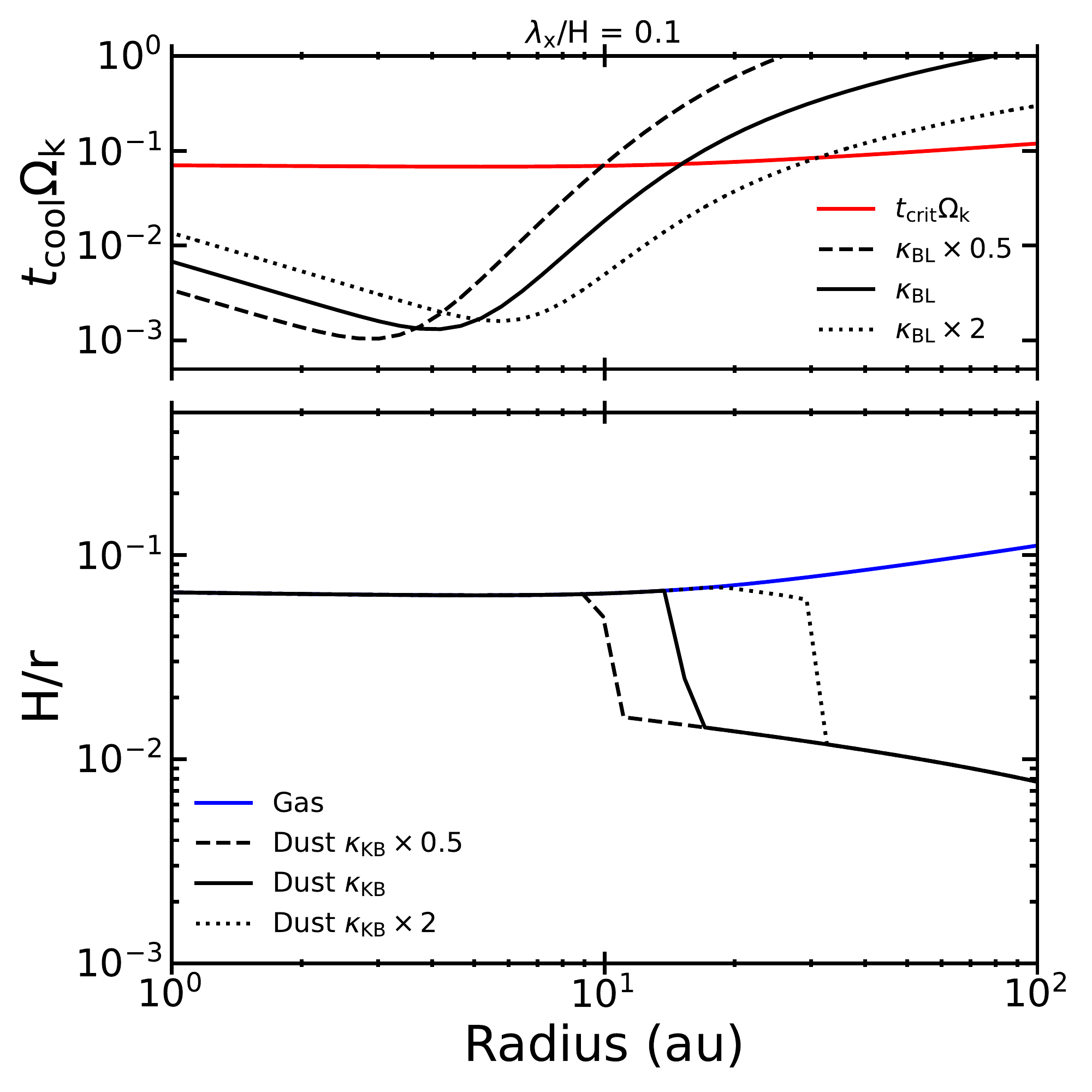}
    \caption{{Top: The dimensionless cooling time $t_\mathrm{cool}\Omega_\mathrm{k}$, assuming a VSI wavelength mode of $\lambda/H = 0.1$. Below a critical cooling time (red), the disk cools quickly enough that the VSI can become active. The sensitivity of the VSI-active region on opacity ($\kappa_{\rm{KB}}$) is illustrated by increasing (dashed) or decreasing (dotted) the opacity by a factor two. Bottom: The effect of VSI turbulence in the inner disk on the scale-height of millimeter-sized particles. The gas scale height is shown in blue, while the mm-dust scale height is shown in black for our nominal opacity. Also here the sensitivity on the opacity is shown with dashed and dotted curves.}}
    \label{fig:beta_Hd_lambda01}
\end{figure}

{
Figure~\ref{fig:beta_Hd_lambda01} shows  the relation between the midplane cooling time and the critical cooling time for the VSI to operate in more detail. Here we use standard assumptions that follow \citet{Lin2015} and a VSI unstable wavelength of $\lambda_x/H=0.1$ (see Appendix \ref{app:VSI} for details and a further parameter exploration).
The requirement for sufficiently short cooling times is typically not met in the very outer optically-thin parts of the disks ($\gtrsim 50$\,au).
However, the exact location of this outer boundary of the VSI active region depends sensitively on the dust opacity, as $t_{\rm cool} \propto \kappa^{-1}$ \citep{Lin2015}. 
For our disk model we find that the VSI cooling criterion is satisfied out to $20$\,au, using a nominal Rosseland mean dust opacities \citep{BellLin1994}.
A factor two change in opacity can lead to a factor two change in outer boundary radius.
}

%


{
Furthermore, fig.~\ref{fig:beta_Hd_lambda01} shows the resulting particle scale height of mm-sized grains, when using a conservative vertical turbulent stirring rate of $\alpha_{\rm z} = 2.5 \times 10^{-3}$ in the VSI-prone region. This vertical stirring $\alpha_{\rm z}$ is based on global numerical simulations of VSI-unstable disks \citep{Flock2017}. 
The vertical stirring by the VSI gas motions puffs up the pebble layer, such that millimeter size grains are elevated up to the gas scale height over nearly the entirety of the VSI-prone region.
Outside this radius, we have here assumed lower vertical particle stirring, $\alpha_{\rm z} = 1\times 10^{-4}$, more in line with outer disk non-ideal MHD particle diffusion \citep{Riols2018} and the observed low particle scale heights at wide orbits in other protoplanetary disks \citep{Pinte2016,Villenave2022}.
}

\section{Discussion}

\subsection{Implications for planet formation}

\begin{figure}
    \centering
    \includegraphics[width = \hsize, trim={4cm 0cm 3cm 0cm}, clip]{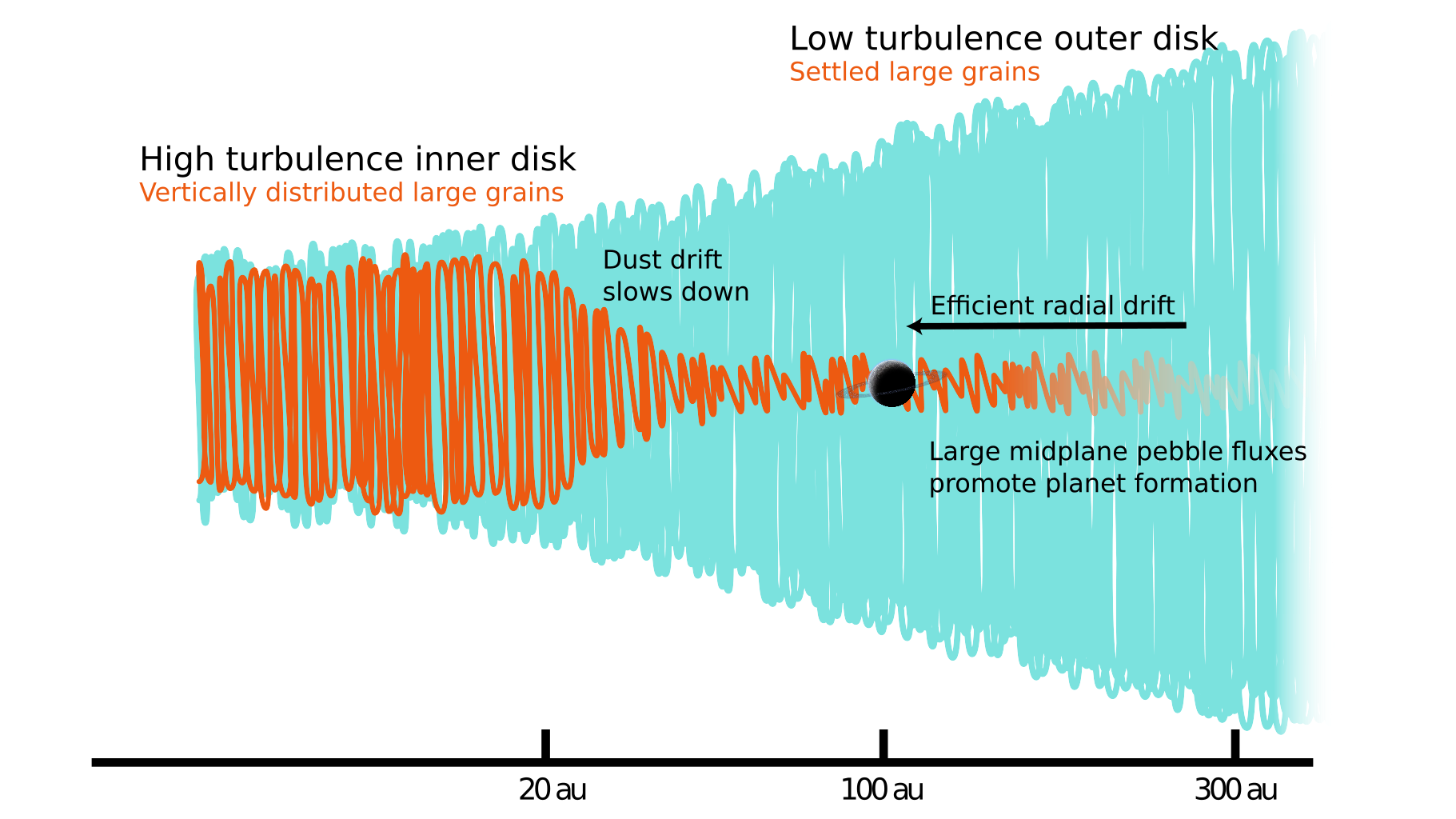}
    \caption{Schematic of the dust (orange) and gas (blue) distribution in the IM Lup disk. A planet image has been put in the approximate position of a proposed proto-planet around 100 au \citep{Pinte2020, Verrios2022}. }
    \label{fig:schematic}
\end{figure}


The observations of IM Lup imply a split between two regions of the disk, an inner region ($\lesssim$ 20 au) where dust is piled-up and vertically extended distributed, and an outer disk ($\gtrsim$ 20 au) where dust is efficiently being transported in through radial drift (see Fig.~\ref{fig:schematic}). 

{Efficient drift in the outer disk implies relatively high Stokes numbers ($>$ 0.01) \citep{Weidenschilling1977, Birnstiel2010} and high St/$\alpha$ \citep[e.g.][]{Birnstiel2012}, which should lead to well settled grains \citep[][]{Johansen2014}.} The inner disk dust pile-up implies that at least $110$ $M_\oplus$, and more likely $\gg 110$ $M_\oplus$ has been transported from the outer disk into the inner 20 au within the 1 Myr age of the IM Lup disk \citep{Mawet2012}. This scenario is ideal for the formation of giant planet cores at larger ($>20$ au) radii \citep{Bitsch2019, Johansen2019}. 

In contrast, the high pebble scale-heights, implying low $\mathrm{St}$ or high $\alpha$, required to fit the observations will have an adverse effect on the efficiency of pebble accretion, quenching giant planet formation in the majority of the inner 20 au region \citep{Lambrechts2014}. The Stokes numbers for the dust in the inner disk must be small to impact the CO emission. For example, 1 mm particles at 20 au have a $\mathrm{St} = 6\times10^{-4}$. This implies that even with the large dust surface densities, and high metallicity in the inner 20 au, the streaming instability might not be triggered \citep{Carrera2015, Yang2017, Li2021}.  



If a stage like this is common for massive proto-planetary disks, it could explain the formation of planets at large radii invoked to explain dust and gas structures in a variety of older proto-planetary disks  \citep[e.g.][]{Zhang2018, Pinte2022}. In fact, IM Lup is proposed to have a forming planet out at 117 au  \citep[e.g.][]{Zhang2018}. An inner disk with high turbulence and small pebbles as a result of strong pebble drift would also predicts that giant planet formation due to pebble-accretion is strongly suppressed in the inner disk. 

Formation of giant planets this far out in the disk would have massive impact on the composition of the material that they would accrete. Especially the pebbles that are forming the core would bring in large amount of very volatile ices like \ce{CO}, \ce{N2}, \ce{Ar}, \ce{Kr}, and \ce{Xe}, which would be absent from the pebble ice in the warmer inner regions. Enhancements of these species have been observed in the atmosphere of Jupiter in the solar system, implying that Jupiter formed far from its current location and that the early solar system might have gone through a state similar to what we see IM Lup in now \citep{Oberg2019, Bosman2019Jup}.  

Interactions between (forming) planets and their natal disks is expected to cause significant migration of planets. Planets forming at large radii ($>$ 50 au) can end up very close to the star. Planets formed far out can thus be the progenitors of some or all of the giant planets on close orbits we see now. Efficient migration is also required to match the proto-planetary disk mass and size distribution with the observed giant exoplanet distribution  \citep{Mulders2021}. 

\subsection{Alternative scenarios}
A drop in CO isotopologue flux can have other causes than the dust pile-up proposed here. \citet{Zhang2021MAPS} modeled the CO flux with very low CO abundances in the inner disk. {However, derivations from the $J=$2--1 and $J=$1--0 \ce{C^{18}O} lines  disagree on the CO abundance by an order of magnitude. As both these lines come from the same isotopologue, this cannot be reconciled with any chemical explanation. However, if the dust is absorbing line emission then the wavelength dependent opacity of the dust can naturally explain why between the 1.3 mm $J=$2--1 and 3 mm $J=$1--0 \ce{C^{18}O} lines are differently impacted.} The radial profiles by \cite{Law2021aMAPS} show, however that all robustly detected lines{, across all observed species} share the inner disk depression. This is not naturally explained by a low CO abundance, {but is expected from a puffed up inner disk pile-up.}

A reduced gas surface density in the inner disk would be able to explain the decrease line flux. The clear drop in the $J=$2--1 \ce{^{13}CO} line would then imply that this line becomes optically thin. At a conservative inner disk CO abundance of $10^{-5}$ w.r.t. \ce{H2}, this implies a gas surface density $< 0.06$ g cm${-2}$, or a gas surface density decrease of at least 3 orders of magnitude \citep[e.g.][]{Bosman2021bMAPS}. This is inconsistent with the high gas accretion rate $\sim 10^{-8}$ M$_\odot$ yr$^{-1}$ measured for IM Lup \citep{Alcala2017}. 
We therefore conclude that the inner disk dust pile-up is consistent with a larger range of observations than alternative scenarios.

\subsection{Finding more drift heavy disks}
The leaves the question on how common the formation of a dust pile-up with vertically extended dust is. If it is the VSI that is triggered by strong dust drift and subsequent inner disk pile-up, the process might be universal to disks with large dust drift rates. 
To test this however, one needs to find other disks in the same state as IM Lup. Analysing the CO isotopologue emission in the inner disk would be a good way of determining inner disk dust pile-ups. However, observations with the sensitivity and resolution as those available for IM Lup are too time consuming to do for a large sample of disks. As such some pre-selection is required.

In multi-wavelength continuum observations, the inner disk looks different from the outer disk, with the inner disk having a flatter spectral slope than the outer disk. This could be caused by the dust pile-up, increasing the optical depth, or the higher turbulence, changing the grain size distribution and thus optical properties  \citep{Sierra2021MAPS}. As such this could be a marker of extreme drift. 

Episodes of extreme drift are short lived, constrained by the dust mass reservoir of the disk. As such a search should be focused towards young disks, in particular those still embedded in their natal envelope, known as Class I disks. Observations probing the inner $\sim 30$~au regions are rare and a direct counter part of the IM Lup data does not exists for any of these sources. However, \citet{Harsono2020} aimed to detect water vapor originating from the inner regions of these Class I sources. Contrary to expectations, water vapor was not detected. If the inner disks of these Class I objects is shaped by the same processes as IM Lup, then the abundant, lofted dust would suppress the water emission from the inner 20 au, similar to the suppression in the inner disk \ce{C^{18}O} emission from IM Lup. 

Finally, IM Lup stands out in another inner disk tracer, its mid-infrared spectrum. It is one of only a handful of sources that shows just \ce{CO2} without the \ce{H2O}, \ce{HCN} and \ce{C2H2} that are commonly seen in the same part of the spectrum \citep{Salyk2011, Bosman2017}. This could be caused by drift, as well as abundant small grains in the surface of the inner disk and as such could be a signpost of a recent episode of drift and strong stirring in the inner disk. Strong drift is expected to enrich the inner disk with water, which is expected to suppress the abundance of \ce{HCN} and \ce{C2H2} and increase the abundance of \ce{CO2}. Abundant grains in the surface layers limit the region where  mid-infrared molecular lines can be generated. This leaves only the top low density region of the disk above the layer of elevated optically thick dust. In this region the density is too low to excite any of the infrared transitions collisionally. The 15$\mu$m bending mode of \ce{CO2} can be still be excited by infrared continuum photons and would thus still be observable  \citep{Bosman2017}. If this is indeed the case then the bending mode of water at 6.5 $\mu$m might also be visible in emission, even though the pure rotational lines at 12--30 $\mu$m are not observed as these transitions can also be directly excited by continuum photons \citep[][]{Bosman2022water}.

\section{Conclusions}

The IM Lup system poses an interesting case study for planet formation through pebble accretion. The inner 20 au shows a strong enhancement of large dust that requires a continuous or very recent ($<$ 1.0 Myr) massive pebble flux, with pebble drift rates significantly higher than 110 $M_\oplus$ Myr$^{-1}$. These conditions allow for the fast formation of giant planet cores. In the inner 20 au however the dust has to be vertically extended to impact the line emission. A vertically extended dust distribution is predicted to greatly slow down the formation of giant planet cores through pebble accretion. The more settled regions outside 20 au are still conducive to giant planet formation. This would naturally lead to giant planets forming at large radii, such as those that are leaving imprints in dust and gas in many older protoplanetary disks. If the vertically extended dust is a natural consequence of the dust evolution in a drift dominated disk, as the VSI predicts, then it would be expected that all giant planets, including Jupiter, formed their core at radii $>20$ au. 

\acknowledgments 
ADB and EAB acknowledge support from NSF Grant\#1907653 and NASA grant XRP 80NSSC20K0259.
J.A. acknowledges the Swedish Research Council grant (2018-04867, PI A.\,Johansen).
M.L.\, acknowledges funding from the European Research Council (ERC Starting Grant 101041466-EXODOSS).


\appendix

\section{Thermochemical model}
\label{app:therm_model}
To predict the CO isotopologue emission we used the thermochemical code DALI \citep{Bruderer2012, Bruderer2013}. This code allows us to take the physical conditions from the dynamical models, as well as stellar parameters, and calculate temperature and chemical abundance over the 2D model. We can then raytrace the temperature and abundance structure to calculate the emission. The input stellar spectrum used is a stellar model with added UV \citep{Zhang2021MAPS}. The disk structure is based on the IM Lup structure of \citet{Zhang2021MAPS}, disk parameters are given in Table~\ref{tab:DALI_param}.

For the dust optical properties we use the small (0.005-1$\mu$m) and large (0.005-1000$\mu$m) populations from \citet{Zhang2021MAPS} \citep[see also ][]{Birnstiel2018}. The elemental abundances used in our model are shown in Table.~\ref{tab:abu}.  Contrary to \citet{Zhang2021MAPS} we assume a constant CO depletion factor of 100, the value measured at 100-150 au. 
To make a proper comparison with the data \citep{Law2021aMAPS}, the image cubes from DALI are post-processed. The individual channels in the cube are convolved with a circular gaussian with a FWHM of 0.15'', before the channels are summed and a integrated intensity map is made. From this map a radial profile is extracted using a 30 degree wedge around the semi-major axes of the disk.  

\begin{figure}
    \centering

    \includegraphics[width = \hsize]{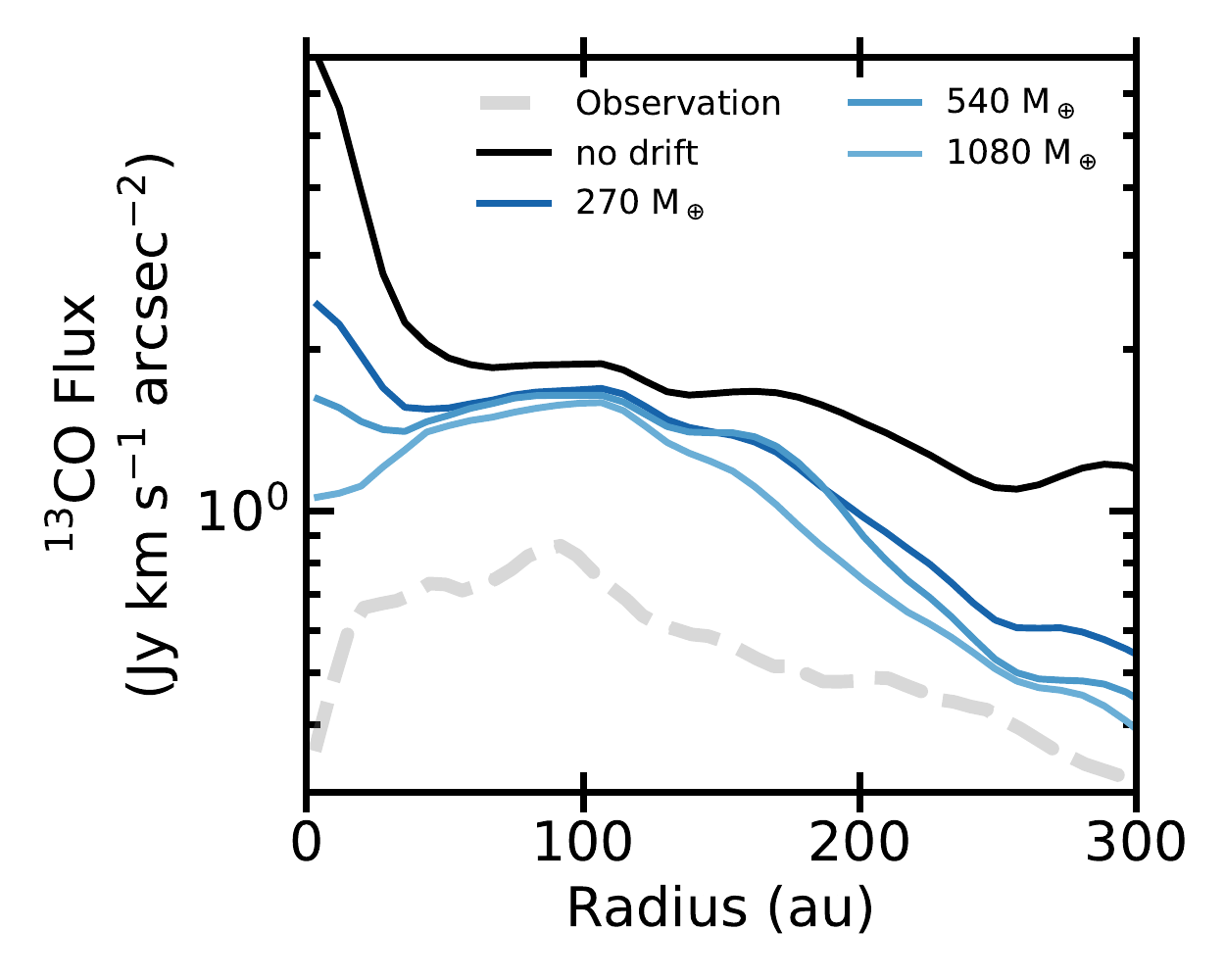}
    \caption{Same as Fig.~\ref{fig:C18O_rad} but for \ce{^{13}CO} instead of \ce{C^{18}O}. Again, $>$540 M$_\oplus$ is required to create an inner disk depression. 
    }
    \label{fig:13CO_rad}
\end{figure}

We ray-traced the $J$=2--1 transitions of \ce{^{13}CO} and \ce{C^{18}O}. The \ce{C^{18}O} traces the deepest into the disk of these three lines and is thus the most sensitive to dust in the inner disk (see Fig.~\ref{fig:C18O_rad}). \ce{^{13}CO} traces higher up near the disk surface. These radial profiles are shown in Fig.~\ref{fig:13CO_rad}. While the \ce{^{13}CO} is over predicted by the model. The behavior of the \ce{^{13}CO} still mirrors that of the \ce{C^{18}O}, in the fact that, for the models with $>450 M_\oplus$ within 20-30 au, a dip in the emission profile can be seen. As such the \ce{^{13}CO} emission supports our finding out of the \ce{C^{18}O} emission.

\begin{table}[]
    \centering
    \caption{Thermochemical modeling parameters}
    \begin{tabular}{l c c}
    \hline
    \hline
    parameter & value & explanation \\
    \hline
        $M_\star$ & 1.1 $M_\odot$ & Stellar mass\\
        $L_\star$ & 2.57 $L_\odot$ & Stellar luminosity \\
        $M_\mathrm{gas}$ & 0.2 $M_\odot$ & disk gas mass \\
        $M_\mathrm{dust, large}$ & 0.002 $M_\odot$ & large dust mass \\
        $M_\mathrm{dust, small}$ & $2\times 10^{-4}$ $M_\odot$ & small dust mass \\
        $R_\mathrm{crit}$ & 100 au & critical radius \\
        $\Sigma_c$ & 28.4 g cm$^{-2}$ & gas surface density at $R_c$\\
        $\gamma$ & 1 &  gas surface density slope \\
        $h_c$ & 0.1 & gas scale height at critical radius \\
        $\psi$ & 0.17 & flaring angle \\
        \hline
    \end{tabular}
    
    \label{tab:DALI_param}
\end{table}

\begin{table}[]
    \centering
        \caption{Elemental abundances w.r.t H}
    \begin{tabular}{l|c}
    \hline \hline
        Element & Abundance w.r.t. H \\
    \hline
        H &  1.0 \\
        He & $7.59 \times 10^{-2}$ \\
        C & $1.35 \times 10^{-6}$ \\
        N & $2.14 \times 10^{-5}$ \\
        O& $2.88 \times 10^{-6}$ \\
        Mg& $4.17 \times 10^{-9}$ \\
        Si& $7.94 \times 10^{-8}$ \\
        S& $1.91 \times 10^{-8}$ \\
        Fe& $4.27 \times 10^{-9}$ \\
    \hline
    \end{tabular}

    \label{tab:abu}
\end{table}

\section{Dust evolution model}
\label{app:Dust_evolution_model}
We ran a disk evolution model similar to \citet{Appelgren2020}, which includes disk formation, viscous evolution, and radial drift of dust, assuming different dust sizes. The formation of the disk is modeled starting from the gravitational collapse of an over-dense Bonnor-Ebert sphere  \citep{Bonnor1956, Ebert1957}. During disk formation the largest radius at which material lands on the disk is the maximum centrifugal radius, given by the following equation
\begin{align}
	R_\mathrm{c} = \dfrac{\Omega_0^2 r_\mathrm{cf}\left( t \right)^4}{G M\left( r_\mathrm{cf} \right)} \label{eq:Rc}.
\end{align}
Here, $\Omega_0$ is the solid rotation rate of the cloud core, $r_\mathrm{cf}$ the radius of the outwards expanding collapse front, and $M\left( r_\mathrm{cf} \right)$ the total mass inside the collapse front radius. For a molecular cloud core of a given mass, changing the centrifugal radius effectively sets its angular momentum.  Different values for $R_\mathrm{c}$ therefore result in different disk sizes and masses.

The speed at which dust particles drift depends on their Stokes numbers, with a maximum drift rate at a Stokes number around unity. In the Epstein drag regime the Stokes number is set by the following equation
\begin{align}
    \tau_\mathrm{s} =  \dfrac{\sqrt{2\pi}\rho_\mathrm{d} a_\mathrm{d}}{\Sigma_\mathrm{g}}, \label{eq:stokes}
\end{align}
 where $\rho_\mathrm{d}$ is the material density of the dust particles, $a_\mathrm{p}$ the size of the dust particles, and $\Sigma_\mathrm{g}$ is the gas surface density. 
 
 The mass of the dust disk is determined not only by the mass and angular momentum of the cloud core, but also by the assumed dust-to-gas ratio of the cloud core. {We will keep the dust-to-gas ratio fixed to the nominal ISM value of Z=0.01}. 
 
{We ran the disk evolution model for a grid of centrifugal radii and dust sizes}. The gas disk in IM Lup extends out to $1200$\,au \citep{Zhang2021MAPS}. Because of the large size of the gas disk, the grid of centrifugal radii ranged from $150$\,au to $300$\,au in steps of 50\,au. These large values of the centrifugal radius, together with a viscous $\alpha$ parameter of $\alpha_\nu = 10^{-2}$, ensures that the gas disks is able to expand out to about $1000$\,au within $1$\,Myr, which is the estimated age of IM Lup. We ran the model with fixed particle sizes of $0.1$, ${0.2}$, $0.5$, $1$, and $3$\,mm. 

From this grid of models, the case which best fits IM Lup had a centrifugal radius of 250 au, 1-mm-sized dust grains. The age of this disk when it best matched IM Lup was 0.59\,Myr. This selection was based on the model piling up sufficient dust ($\sim$\,500\,M$_\oplus$) within the inner $20$\,au \citep{Bosman2021bMAPS,Sierra2021MAPS}, while having a gas radius which extends to about 1000\,au, and an age of about 1\,Myr or less when this pile-up occurs. 

\begin{figure*}
    \centering
    \includegraphics[width = \hsize]{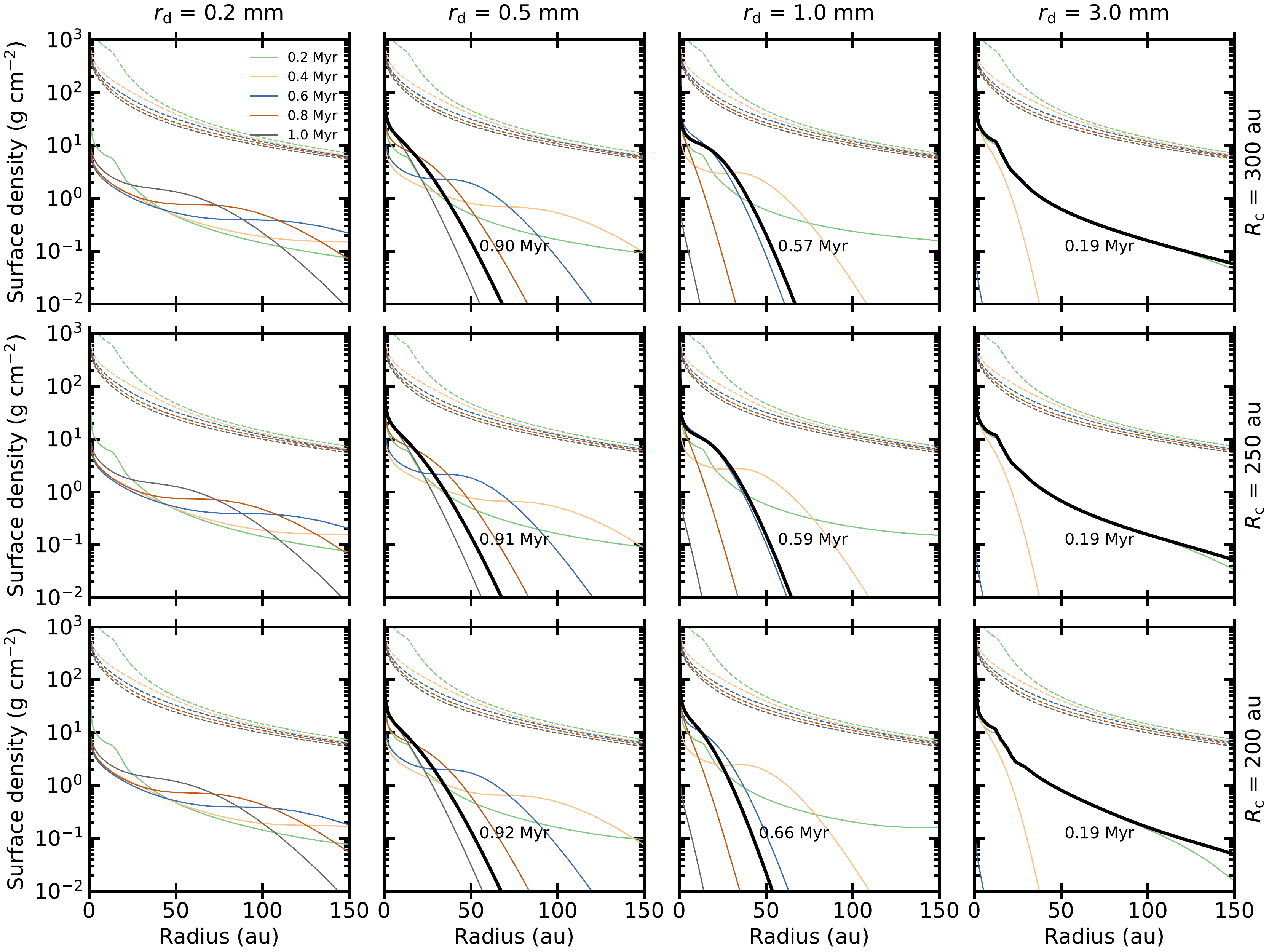}
    \caption{The evolution of the gas (dashed)  and dust (solid) surface density for a selection of models from the grid of parameters explored. Each panel shows a combination of particle size and centrifugal radius. All models shown here used an initial dust-to-gas ratio of $Z=0.01$. The black lines indicate the time, as labeled, at which 500 $M_\oplus$ have piled-up inside 20 au. The model selected as the best fit for IM Lup is shown in the center-right
    panel, with a particle size of 1 mm and a centrifugal radius of 250 au. Changing the centrifugal radius has a minor effect on the evolution of the dust, but it affects the size of the gas disk at the time of the pile-up. Models with particle sizes of 3 mm only reach 500 $M_\oplus$ of dust within 20 au just after the disk has finished forming, and would thus likely represent objects that are still embedded in natal envelope. 
    In the models with dust particles that are 0.2 mm in size, dust drift is too inefficient to pile-up 500 $M_\oplus$ in the inner 20 au at any time.
    }
    \label{fig:surface_dens_dust_mod}
\end{figure*}

To explore the dependency of our preferred model to key parameters we explored variations on the total mass of solids and the radial extent of the inner disk dust pile-up. The resulting surface densities are shown in Fig.~\ref{fig:surface_dens_dust_mod}. The choice of centrifugal radius has a very minor effect on the disk evolution. A smaller centrifugal radius delays the dust pile-up very slightly. Particles sizes smaller than 0.5 mm do not lead to a significant pile-up in the inner disk within 1 Myr (column 1, Figure \ref{fig:surface_dens_dust_mod}). An increased particle size results in earlier pile-ups, such that for the $3$\,mm-sized dust, the pile-up occurs just when the disk has finished forming. These disks would represent object still embedded in their natal envelope and are therefore rejected. Models with 0.5 or 1\,mm-sized dust display similar pile-ups, but the $0.5$\,mm models pile-up about 0.3 Myr later.



We finally verify that the gas accretion rate onto the host star, regulated by the disk mass and chosen $\alpha_nu$ value,  is consistent with the young age of IM Lup. The nominal disk results in a sufficiently high accretion rate of $1.2\times 10^{-7}\ M_\odot/\mathrm{yr}$, which, given uncertainties, moderately exceeds inferred values in IM Lup of $\sim 10^{-8} M_\odot/\mathrm{yr}$ \cite{Alcala2017}.

\section{VSI}
\label{app:VSI}

{
In this Appendix we explore how the radial range of the VSI-prone region depends on various model assumptions, with the aim to demonstrate that in a plausible parameter regime the VSI could be an explanation for the high particle scale heights inferred in the inner disk of IM Lup. We directly follow here the work of \citet{Lin2015}.
}

\begin{figure*}
    \centering
    \includegraphics[width = \hsize]{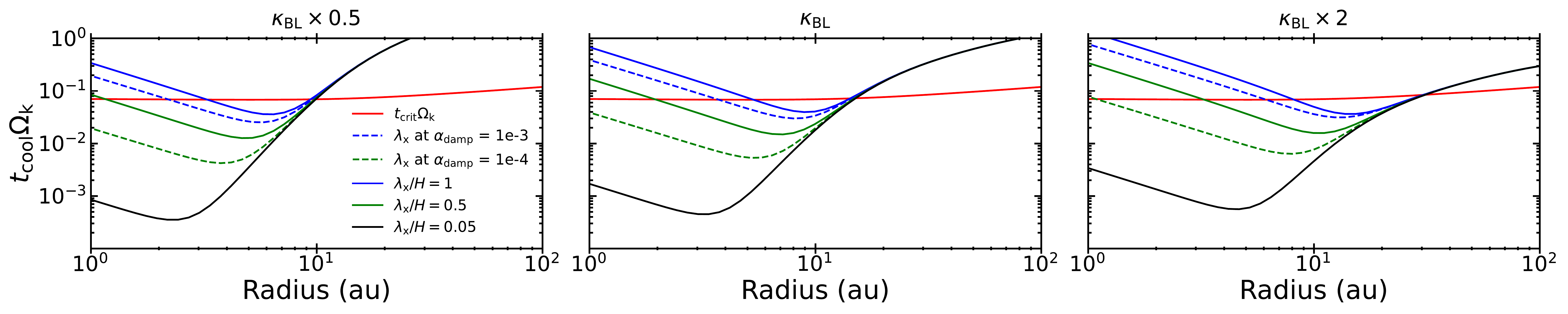}
    \caption{The effect of Rosseland mean opacity ($\kappa_\mathrm{BL}$) and  {VSI wavelength mode ($\lambda_x$)} on the dimensionless cooling time parameter $t_\mathrm{cool}\Omega_\mathrm{k}$. At cooling times below $t_\mathrm{crit}\Omega_\mathrm{k}$ the VSI can become active. 
    An increased opacity increases the cooling time in the inner optically thick regions of the disk and decreases the cooling time in the outer optically thin disk regions. As a result the VSI-prone region shifts outwards in the disk.
    }
    \label{fig:beta}
\end{figure*}

{
As discussed in the main text, our results strongly depend on the assumed dust opacity. 
We therefore show our cooling time results for different Rosseland mean dust opacities, with a temperature dependency following \citet{BellLin1994} and an MRN-dust distribution \citep{Savvidou2020}. 
Then, by reducing, or increasing, the opacity by a factor 2, we illustrate the trend when considering, respectively, sub-solar or super-solar mass fractions of sub-10 $\mu$m grains. 
The opacity-dependency of the cooling times as function of orbital radius can be seen in Fig.\,\ref{fig:beta}.
An important caveat is that the dust distribution and resulting opacities are not well constrained for IM Lup, and may also be different in the inner and outer disk. 
A further exploration of the effects of the dust growth and disk evolution can be found in \citet{Fukuhara2021}, who find the VSI suppressed in sub-solar dust-to-gas environments. 
Future work could thus aim to link the opacity to a modeled particle size distribution and local dust-to-gas ratio.
}

\begin{figure}
    \centering
    \includegraphics[width = \hsize]{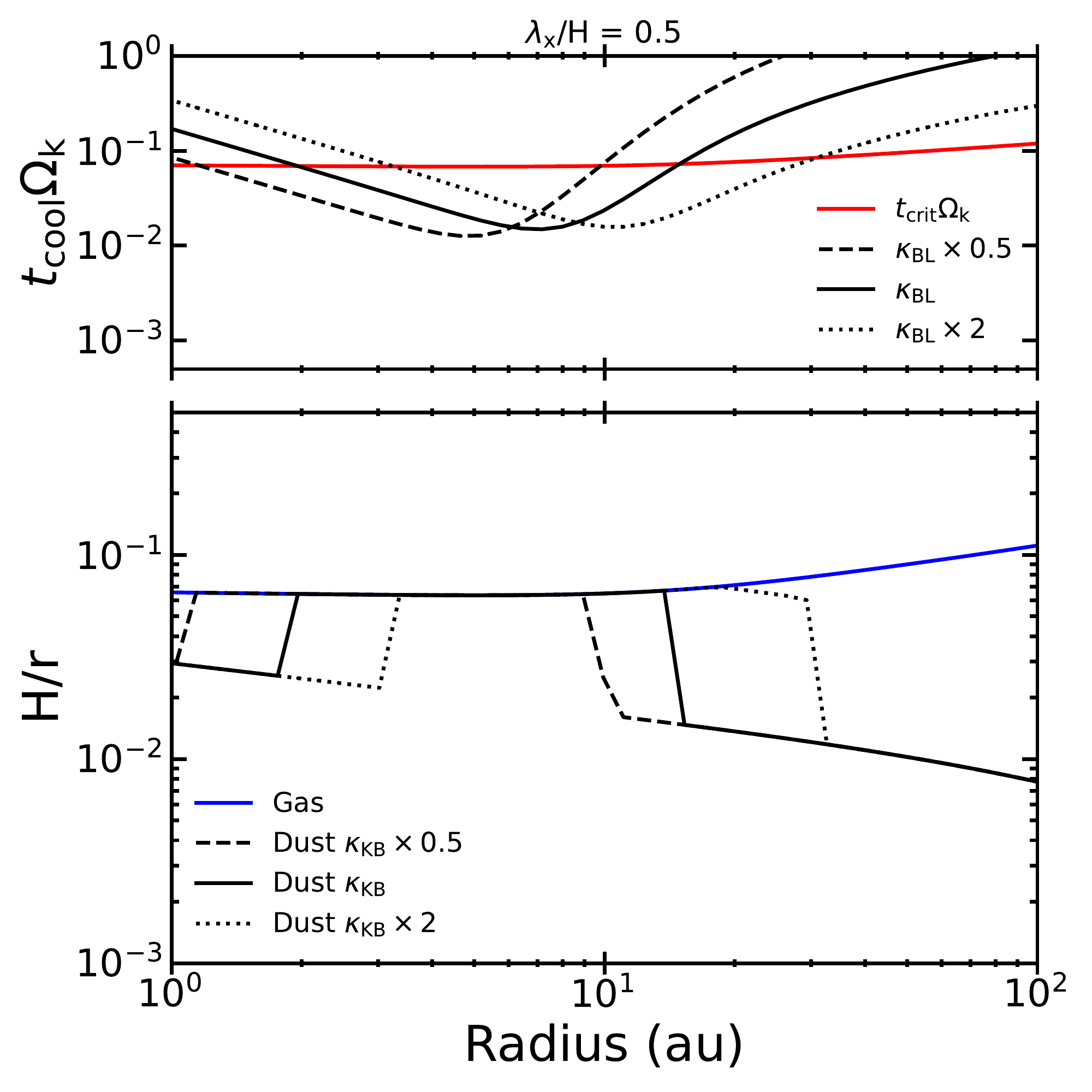}
    \caption{Same as figure \ref{fig:beta_Hd_lambda01}, but assuming a wavelength mode of $\lambda_\mathrm{x} = 0.5\ H_\mathrm{g}$. At longer wavelengths cooling is less efficient in the optically thick regions of the disk. As a result, the inner edge of the VSI-active region moves outwards and mm-sized dust particles remain settled in the innermost region of the disk.}
    \label{fig:beta_Hd_lambda05}
\end{figure}

{
In the inner optically-thick part of the disk, the cooling time depends on the length scale of the fastest growing VSI-mode.  We explore different values, with modes between $\lambda_x/H=0.05,0.5,1$ in Fig.\,\ref{fig:beta}, in line with the typical parameter range explored in VSI studies $\lambda_x/H \sim \mathcal{O}(0.5)$ \citep[e.g.][]{Pfeil2019}.
For the largest scale VSI modes, comparable to the gas scale height, we find the VSI to be nearly fully suppressed, with the exception of a small region around $10$\,AU. 
We also illustrate how the VSI-region with high particle scale height is reduced for the $\lambda_x/H=0.5$-case (Fig.\,\ref{fig:beta_Hd_lambda05}), compared to the $\lambda_x/H=0.1$-case in the main text (Fig.\,\ref{fig:beta_Hd_lambda01}). 
Fig.\,\ref{fig:beta_Hd_lambda05} illustrates that the inner edge of the VSI region is now located withing the inner few AU of the disk. Such a close-in inner VSI edge in the particle scale height would not be observable with the CO observations presented here. 
}

{In practice, it is not clear how to determine the fastest growing VSI mode for the IM lup disk model, as other sources of turbulence could damp small-scale modes.
\citet{Lin2015} provide a heuristic argument to determine a minimal growth scale by requiring the growthrate to exceed the viscous timescale 
$t_{\rm visc} \approx \lambda^2/\alpha_{\rm damp} c_s H$ on that scale. 
However, the appropriate value for $\alpha_{\rm damp}$ remains uncertain, as $\alpha_{\rm damp}$ is not necessarily equal to the vertical particle $\alpha_z$ or the ad-hoc $\alpha_\nu$ to evolve the disk in time (Appendix \ref{app:Dust_evolution_model}). 
Indeed, simulations of VSI turbulence with the streaming instability \citep{Schafer2022}, or disk turbulence under non-ideal MHD conditions \citep{Cui2020}, show a complex interplay. With these caveats in mind, Fig.\,\ref{fig:beta} shows results for a viscous cut-off, with $\alpha_{\rm damp}=10^{-4}$ and $\alpha_{\rm damp}=10^{-3}$ (dashed lines). Larger vales of $\alpha_{\rm damp}$ would drive scales towards the local gas scale height and suppress the VSI on global scales.
}

%
%
%
%

\bibliographystyle{aasjournal}
\bibliography{Lit_list}

\end{document}